\documentclass{WileyMSP-template}

\begin{document}

\pagestyle{fancy}

\title{Multilayer Q-BIC-like Optical Filters with High Throughput Direct-Write Multilayer Lithography}

\maketitle


\author{A. Bilgehan Baspinar}
\author{Phillippe Pearson}
\author{Andrei Faraon*}


\dedication{}

\begin{affiliations}
A. B. Baspinar, P. Pearson\\
T. J. Watson Laboratory of Applied Physics and Kavli Nanoscience Institute, California Institute of Technology, 1200 E. California Blvd., Pasadena, CA 91125, USA\\
abaspina@caltech.edu

Prof. A. Faraon\\
T. J. Watson Laboratory of Applied Physics and Kavli Nanoscience Institute, California Institute of Technology, 1200 E. California Blvd., Pasadena, CA 91125, USA\\
faraon@caltech.edu

\end{affiliations}




\begin{abstract}
Multilayer metasurfaces provide substantially greater spectral design freedom than single-layer devices, yet their implementation in the visible and near-infrared remains limited by the complexity, cost, and low throughput of conventional nanofabrication. Here, we establish a recently proposed direct-write electron-beam lithography approach as a high-throughput fabrication platform for multilayer resonant metasurfaces, based on an antimony precursor that decomposes in situ into high-index Sb$_2$S$_3$. This method eliminates deposition–etch cycles and reduces each layer to only two fabrication steps, enabling efficient realization of multilayer architectures. Using this platform, we demonstrate multilayer q-BIC–derived metasurfaces with independently tunable resonance wavelengths and linewidths, allowing the construction of compact multi-resonant filters with spectrally decoupled layers. We experimentally demonstrate three-layer devices supporting three resonances and show independent control of resonance wavelength and Q factor across layers. Leveraging this capability, we generate decorrelated filter arrays for compressive sensing and hyperspectral reconstruction, achieving sets of 9 and 36 filters with average absolute Pearson correlation coefficients of 0.11 and 0.21, surpassing prior metasurface and photonic-crystal implementations. These results establish a practical route toward scalable multilayer resonant metasurfaces for spectral filtering, on-chip spectroscopy, and computational imaging.
\end{abstract}



\section{Introduction}
Optical metasurfaces are planar optical devices composed of arrays of subwavelength scatterers that enable precise spatial control of light. While single-layer metasurfaces have demonstrated functionality comparable to that of conventional optical components and more in a compact form factor, more complex and multifunctional objectives such as achromatic responses \cite{multilayerlit_achrom_avayu2017,multilayerlit_achrom_jordaan2025,multilayerlit_achrom_zhang2025,multilayerlit_achrom_zhao2012,multilayerlit_achrom_zhou2018}, optical sorting \cite{multilayerlit_sort_xu,multilayerlit_sort_chen2024,multilayerlit_sort_khan2025,multilayerlit_sort_roberts2023,multilayerlit_sort_mansouree2020}, and optical computation \cite{multilayerlit_comp_li2024,multilayerlit_comp_lin2018,multilayerlit_comp_luo2022,multilayerlit_comp_soma2025,multilayerlit_comp_zhou2024} require additional degrees of freedom in the design and optimization space. Volumetric and multilayer metamaterials provide such expanded design freedom \cite{multilayerlit_chang2025}. However, this design space remains largely unexplored in the visible and near-infrared (NIR) spectral regimes, primarily due to the challenges associated with fabricating high-resolution three-dimensional and multilayer nanostructures. Conventional nanofabrication workflows which typically involve repeated cycles of deposition, electron-beam or photolithography, etching, and planarization, become increasingly complex, costly, and low throughput when extended to multilayer metasurface architectures. Several commercially available fabrication platforms aim to reduce process complexity for multilayer structures, including direct laser writing techniques such as two-photon lithography (TPL), as well as imprint-based approaches. Nevertheless, these methods either lack the resolution required for operation in the visible and NIR regimes or restrict the device architecture to a limited set of polymers with relatively low refractive indices for optical applications. As a result, despite their promise, multilayer metasurfaces remain far less explored than single-layer designs, primarily due to the lack of a scalable, high-yield nanofabrication platform.

Thermal decomposition of metal–organic molecular precursors has been widely employed to form metal chalcogenide films via post-deposition annealing, particularly in the context of solar cell research \cite{han2020,han2021,chemical_wang2012,chemical_wang2015,wang2017}. To the best of our knowledge, this chemistry was first adapted for direct-write patterning in Ref. \cite{Wang2020}, where nanostructures were realized using electron-beam lithography (EBL) and photolithography. More recently, this approach also has been extended to the realization of Fresnel zone plates and metalenses \cite{wangGrayscaleElectronBeam2024}. In this work, we establish this fabrication method as a high-throughput direct-write platform for multilayer metasurfaces, as well as resonant metasurfaces, using an antimony-based organic precursor that decomposes in situ into antimony (III) sulfide (Sb$_2$S$_3$). By leveraging this platform, a conventional multilayer nanofabrication process can be simplified to only two steps per layer: direct EBL writing of the spin-coated antimony-based precursor and spin-coating of the planarization layer.

Quasi-bound states in the continuum (q-BIC) metasurfaces have emerged as a powerful platform for realizing high quality factor ($Q$) optical resonances in subwavelength photonic structures. By perturbing symmetry-protected BIC modes, these metasurfaces enable strong light–matter interactions with precisely engineered radiative coupling, allowing narrow linewidths and large field enhancements to be achieved in compact, planar geometries \cite{forouzmand2017,koshelev2018,joseph2021,campione2016,kanyang2022,yang2025,kuhner2023}. Owing to their high spectral selectivity and geometrically tunable Q factors, q-BIC metasurfaces are promising candidates for constructing filter arrays with directly controllable spectral features for hyperspectral imaging and spectral reconstruction. In such systems, wavelength information is encoded through spectrally decorrelated filters, and the ability to generate low-correlation spectral responses directly determines the reconstruction performance and achievable device compactness \cite{gao2022,liu2021,wang2023,xiong2022,craig2018,wang2019,mcclung2020}. In this work, we design a multilayer q-BIC–derived optical filter architecture with independently tunable spectral features and propose its application to spectral reconstruction and compressive sensing within a compact multilayer metasurface platform.

\section{Results and Discussion}
\subsection{Fabrication Flow}
The fabrication approach employed in this work is a direct-write lithography method in which the exposure-sensitive medium is not a conventional resist, but rather the optical material itself. Direct patterning has been explored as an efficient route for transferring nanoscale features; however, prior demonstrations have largely focused on metals and metal oxides for electronic applications \cite{metaloxidewrite_alexe2000,metaloxidewrite_bhuvana2008,metaloxidewrite_chaker2021,metaloxidewrite_jeong2016,metaloxidewrite_rim2014,metaloxidewrite_zarzar2012}, with comparatively limited exploration in optical devices. Thermal decomposition of metal–organic molecular precursors, such as metal–BDCA (butyldithiocarbamic acid) complexes, has been widely employed in solar cell research, where metal chalcogenide films are formed via annealing-induced decomposition of the metal-organic complexes \cite{chemical_wang2012,webber2013,brutchey2015,webber2014,chemical_wang2015}. In particular, antimony-BDCA precursor solutions have been used to produce Sb$_2$S$_3$ thin films through thermal decomposition for photovoltaic applications \cite{wang2017,han2021,han2020}. More recently, it was demonstrated that Sb$_2$S$_3$, along with other metal/ semiconductor chalcogenides, can be directly patterned through lithography-induced chemical decomposition of metal/ semiconductor–organic precursors \cite{Wang2020}. We focus on Sb$_2$S$_3$, selected for its optical transparency and high refractive index in the visible and NIR spectral regimes \cite{sb2s3refindex_delaney2020,gutierrez2022,2020}, and establish the Sb$_2$S$_3$ direct-write as a fast fabrication approach for multilayer metasurfaces.

This fast multilayer lithography process is based on a solution-processed antimony-BDCA precursor that decomposes into Sb$_2$S$_3$ and organic byproducts upon exposure during electron-beam lithography (EBL). Once synthesized, the precursor remains stable for several weeks and can be readily used by spin-coating the liquid solution onto the substrate and directly patterning it into Sb$_2$S$_3$ (Reaction 3 in Supporting Information, Precursor Solution Synthesis). In the exposed regions, the precursor undergoes in situ conversion into the target chalcogenide material, while the unexposed precursor and organic decomposition byproducts are removed during development by immersing in isopropanol (IPA) (see the details in Fabrication Details section of Supporting Information). For planarization, hydrogen silsesquioxane (HSQ) is employed. HSQ is a negative-tone electron-beam resist that also functions as a spin-on glass owing to the similarity of its chemical composition to that of silica (SiO$_2$). It is a mechanically robust material with a low refractive index, providing a high refractive-index contrast with the Sb$_2$S$_3$ nanostructures. Following planarization, an additional layer of the precursor can be spin-coated, and the direct-write process can be repeated using the same fabrication steps.

\begin{figure}
\centering
\includegraphics[width=\linewidth]{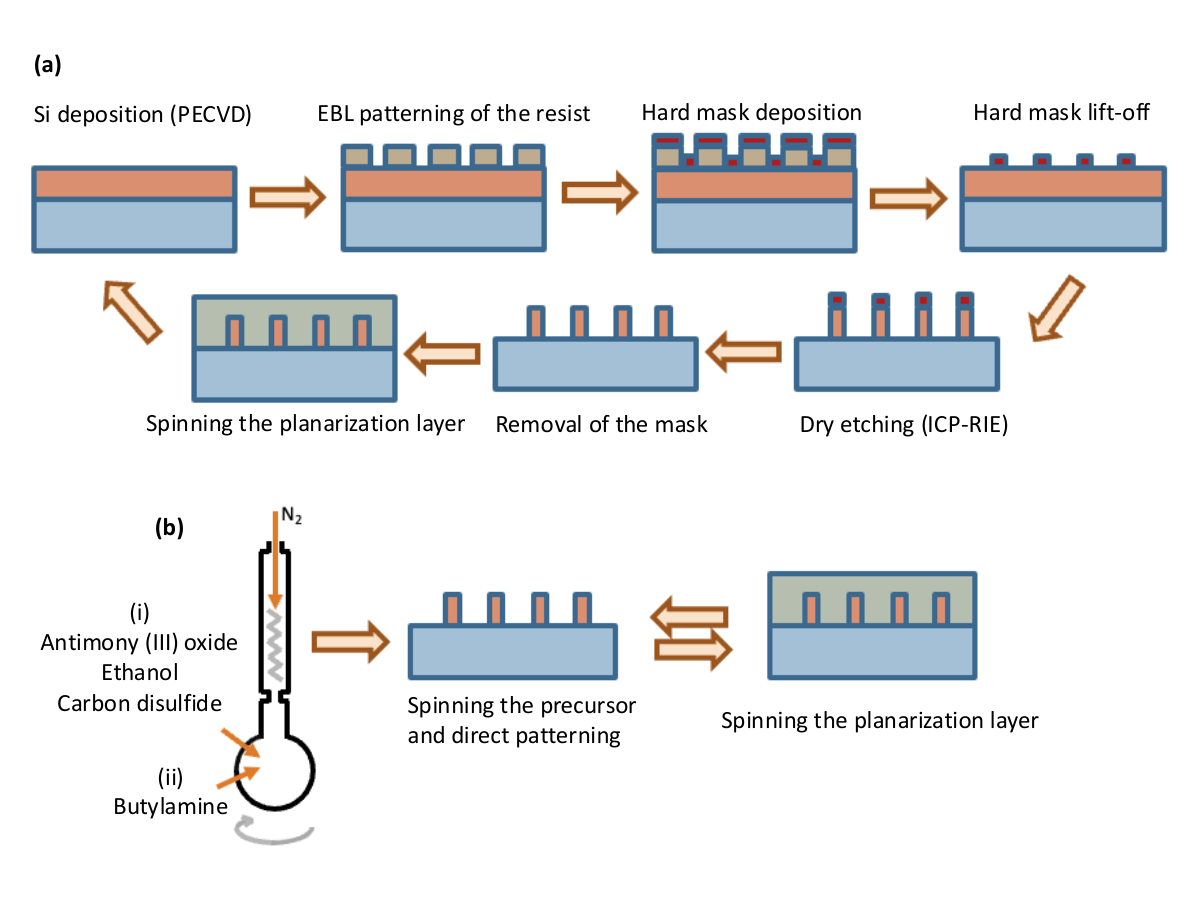}
\caption{(a) An example flow of standard nanofabrication processes to fabricate a multilayer metasurface. (b)Proposed Sb$_2$S$_3$ direct-write process for multilayer metasurfaces.}
\label{fig:fig_fab}
\end{figure}

This approach eliminates both the material deposition and etching steps for each layer, thereby significantly accelerating the overall fabrication process. For comparison, a conventional multilayer EBL process employing materials such as amorphous silicon (a-Si) and spin-on glass typically involves: (1) a-Si deposition using plasma-enhanced chemical vapor deposition (PECVD); (2) electron-beam resist spin-coating; (3) EBL exposure; (4) development; (5) hard-mask deposition; (6) hard-mask lift-off; (7) plasma etching; (8) mask removal; and (9) spin-coating of the planarization layer, such as a spin-on glass or a polymer. These steps must be repeated from step (1) for each additional layer (Figure \ref{fig:fig_fab}.a).
In contrast, the multilayer process introduced here consists of only four steps per layer: (1) spin-coating of the precursor; (2) EBL exposure; (3) development and annealing; and (4) spin-coating of a planarization layer, after which the process is repeated from step (1) for subsequent layers (Figure \ref{fig:fig_fab}.b). As a result, three time-intensive and costly steps, PECVD, etching, and material deposition, are completely eliminated while the interlayer alignment accuracy is still ultimately limited by the lithographic resolution (below 10 nm).

\subsection{Device Design and Experimental Results}

Compressive sensing seeks to recover an unknown signal from a small number of linear measurements, often obtained through randomized or weakly correlated sensing functions \cite{donoho2006}. Because the number of collected measurements is usually much smaller than the signal’s dimensionality, the reconstruction problem becomes an underdetermined linear inversion. The quality of the reconstructed signal depends critically on the degree of randomness and mutual incoherence among the basis functions. In the context of spectral reconstruction, this requirement motivates the use of basis elements with diverse spectral responses, encompassing both broadband and narrowband features, in order to enable high-resolution and robust reconstruction performance.

Leveraging our high-throughput direct-write multilayer lithography platform, we design multilayer, multi-resonant optical filters in which each layer comprises q-BIC–derived metasurfaces with distinct resonance wavelengths and $Q$ factors. The patterned Sb$_2$S$_3$ layers are separated by spacer layers that are sufficiently thick to ensure minimal interlayer coupling. As a result, an $N$-layer device provides $2N$ independent degrees of freedom for spectral control, since both the resonance wavelength and the $Q$ factor of each layer can be independently engineered. This expanded design space is particularly advantageous for spectral reconstruction applications. In the following section, we construct a synthetic set of optical filters with low pairwise spectral correlations to quantitatively demonstrate the effectiveness of these high-throughput multilayer optical filters for compressive sensing and reconstruction.

\begin{figure}
\centering
\includegraphics[width=\linewidth]{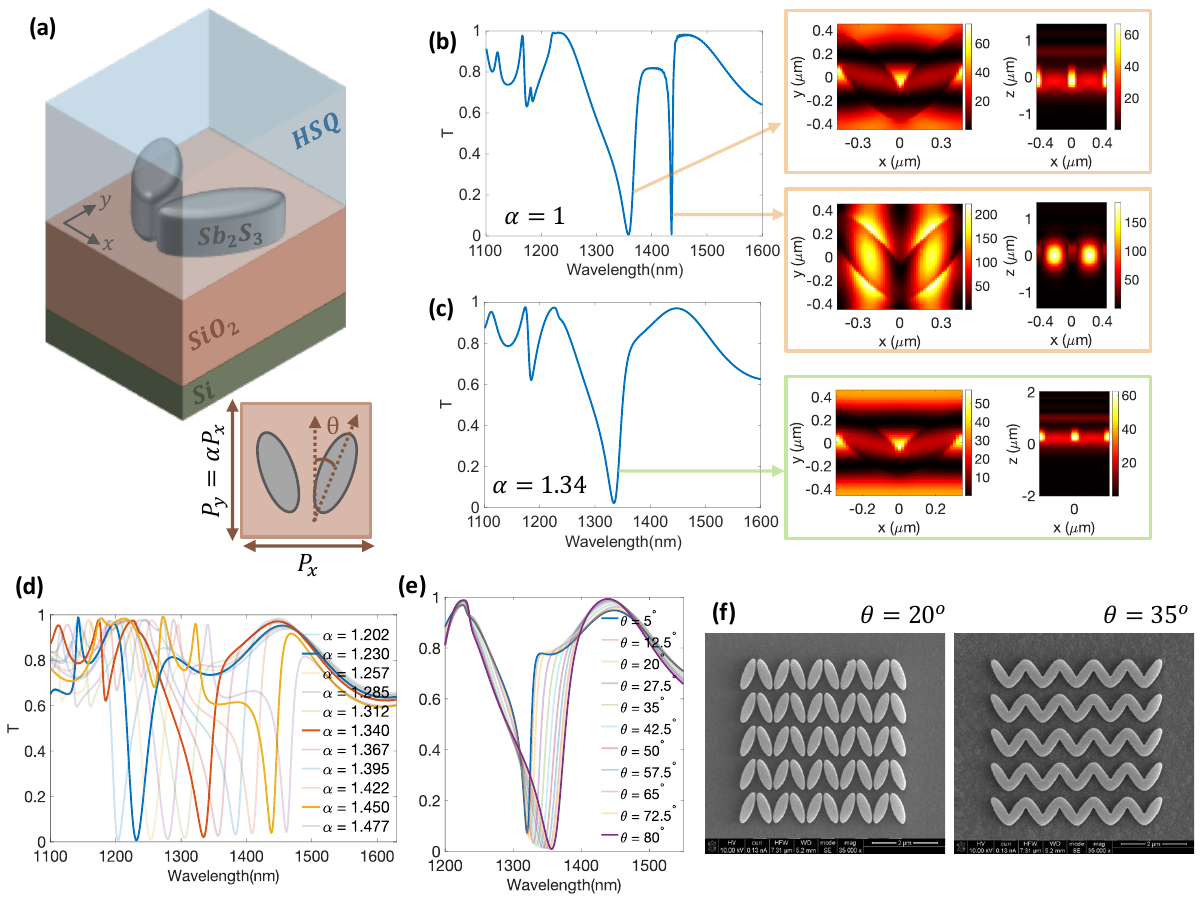}
\caption{(a)The unit cell of the metasurface made of two Sb$_2$S$_3$ tilted posts on top of a 1.3 um thick SiO$_2$ spacer on Si substrate and the definition of parameters $\alpha$ and $\theta$. (b)Simulated transmission for the square lattice ($\alpha$=1, $P_y$=$P_x$=887 nm, $\theta$=35$^circ$), and the norm of the electric field for the inter-post (top) and intra-post (bottom) modes given in the orange boxes. (c)Simulated transmission for the rectangular lattice ($\alpha$=1.34, $P_x$=673nm, $P_y$=$\alpha P_x$=902nm, $\theta$=35$^o$), and the norm of the electric field for the inter-post mode given in the green box. (d)Simulated transmission shown for $\theta$=35$^o$ and $\alpha$=1.202, 1230,.., 1.477.The solid lines refer to the parameter configurations used in Figure \ref{fig:fig_3layer_3wvl}.b. (e)The change in the resonance linewidth with $\theta$=5$^o$,12.5$^o$,..,80$^o$. (f)SEM images of the fabricated Sb$_2$S$_3$ elliptical posts for $\theta$=20$^o$ (left) and $\theta$=35$^o$ (right).}
\label{fig:fig_unitcell}
\end{figure}

The fundamental building block of the metasurface design is a rectangular unit cell with lateral dimensions $P_x$ and $P_y$, as illustrated in Figure \ref{fig:fig_unitcell}.a. Each unit cell contains two elliptical Sb$_2$S$_3$ resonators with a height of 400 nm, positioned on a 1300 nm thick SiO$_2$ spacer layer atop a silicon substrate as shown in Figure \ref{fig:fig_unitcell}.a (see details in Optical Design with FDTD section of Supporting Information). The refractive index of Sb$_2$S$_3$ is taken as 2.1 (Chemical and Optical Characterization section in the Supporting Information), and normally incident light is x-polarized. The major axes of the ellipses are rotated by an angle $\theta$ with respect to the normal, forming a zigzag-like arrangement across the array. This tilted-ellipse geometry has been widely employed to realize q-BIC resonances and has proven effective in a range of applications including biomolecular sensing, optofluidic manipulation of nanoparticles via enhanced electromagnetic fields, and molecular fingerprint detection \cite{leitis2019,tittl2018,yang2022,yesilkoy2019}.

In this design, the resonance wavelength is predominantly controlled by the ratio of the lattice constants $P_y$ and $P_x$, defined as $\alpha = P_y / P_x$, while the quality factor $Q$ (or equivalently, the resonance linewidth) is controlled by the tilt angle $\theta$. These resonances can be observed in the transmission spectrum shown in Figure \ref{fig:fig_unitcell}.b and c. Within the explored parameter space, $\alpha$ is varied from 1.147 to 1.587, enabling tuning of the resonance wavelength from 1150 nm to 1550 nm, a subset of which is shown in Figure \ref{fig:fig_unitcell}.d. Independently, the quality factor is tuned from $Q = 15$ to $Q = 82$ by varying the rotation angle $\theta$ between $5^\circ$ and $80^\circ$ (Figure \ref{fig:fig_unitcell}.e). At larger angles, the two elliptical resonators within a unit cell begin to merge (Figure \ref{fig:fig_unitcell}.f); numerical simulations indicate that this geometric transition does not qualitatively alter the overall evolution of the resonance characteristics.

While the tilted-ellipse geometry supports a single q-BIC resonance when the structure is air-clad, this resonance splits into two distinct modes when embedded in the dielectric environment used here. As shown in Figure \ref{fig:fig_unitcell}.b, the first mode is localized primarily in the gap region between the resonators, exhibits stronger coupling to external radiation, and therefore possesses a lower $Q$. The second mode is predominantly confined within the resonators themselves, couples more weakly to out-of-plane radiation channels, and exhibits a higher $Q$. We refer to these modes as the inter-post and intra-post resonances, respectively.

This resonance splitting arises from the modified dielectric environment relative to the air-clad case, which alters the vertical boundary conditions and redistributes the balance between upward and downward radiation channels \cite{hu2024a,sadrieva2017}. As a result, the inter-post branch no longer follows the ideal q-BIC behavior in which radiative loss vanishes as the perturbation angle approaches zero. Instead, the modified index loading introduces a residual radiative pathway that produces a finite linewidth even at $\theta$ = 0$^o$ (hence the "q-BIC derived" terminology). Although the $Q$ still can be controlled via the rotation-induced perturbations, the presence of this loss channel imposes an upper bound, which is advantageous for practical spectral filtering by improving robustness and signal throughput. 

When the lattice is kept square ($\alpha = 1$), both resonances are present within the transmission spectrum (Figure \ref{fig:fig_unitcell}.b). In contrast, when the lattice is elongated along the $y$ direction ($\alpha > 1$), the intra-post resonance undergoes a blue shift and is pushed outside the operational wavelength band, as shown in Figure \ref{fig:fig_unitcell}.c. Retaining a square lattice thereby enables a richer spectral response with multiple resonance dips exhibiting diverse features.
However, numerical simulations indicate that the quality factor of the intra-post resonances can reach values in the range of $Q \approx 450$–$500$ (Figure \ref{fig:fig_3layer_6wvl}.c). Experimentally, we find that such high-$Q$ resonances are not supported by the present material platform and fabrication approach, as the high-$Q$ intra-post resonances predicted by simulation do not appear in the measured transmission spectra (Figure \ref{fig:fig_3layer_6wvl}.c and d).
Despite this limitation, it is still possible to design devices targeting $2N$ resonances in simulation in order to experimentally realize an effective $N$-resonance device, a strategy that will be discussed later in this section. While this design methodology can be extended to an arbitrary number of layers, we focus on three-layer devices for the remainder of this manuscript for clarity and experimental practicality.

\begin{figure}
\centering
\includegraphics[width=\linewidth]{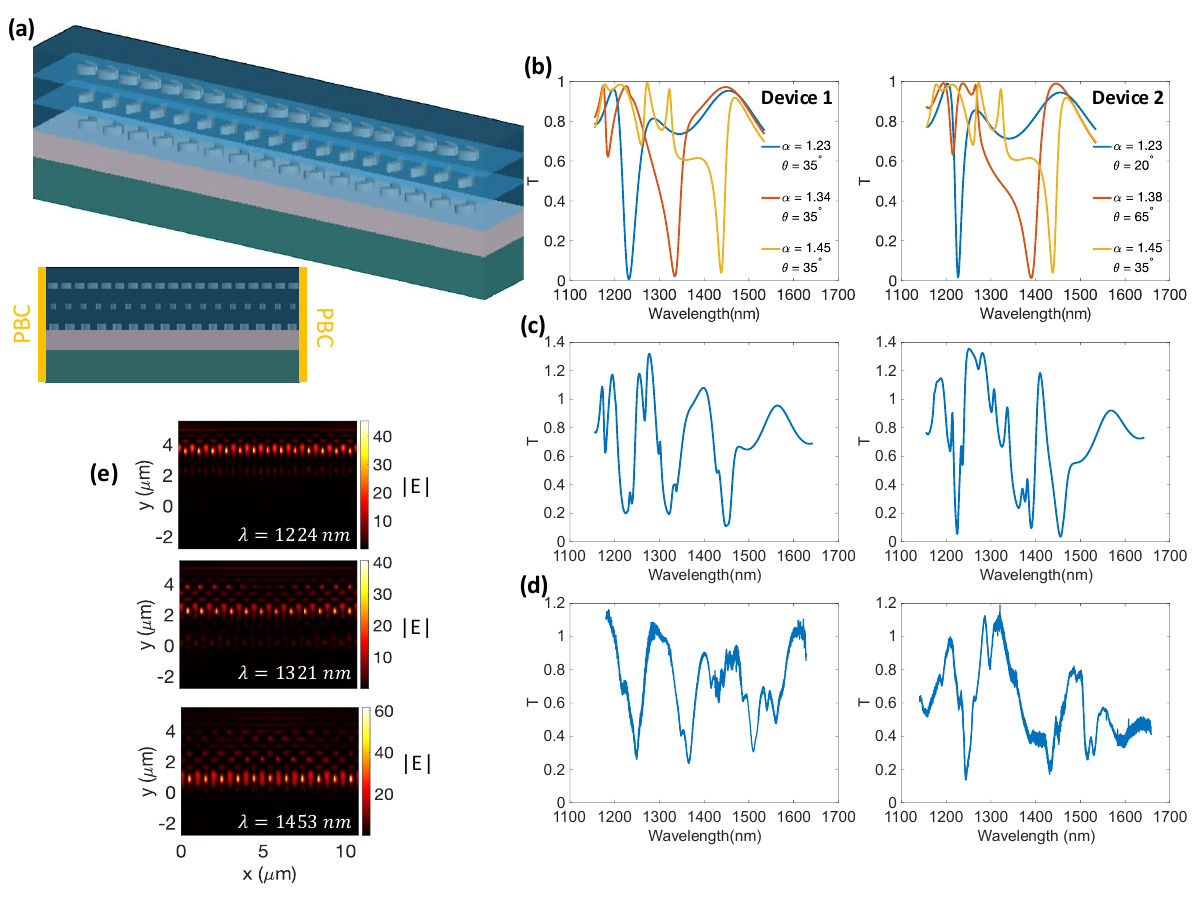}
\caption{(a)The supercell for the rectangular lattice 3 layer device simulation; the side view shows the different number of iterations per layer and the periodic boundary conditions used on sides in FDTD. (b)Single layer simulations for ($\alpha$=1.23,$\theta$=35$^o$), ($\alpha$=1.34,$\theta$=35$^o$), and ($\alpha$=1.45,$\theta$=35$^o$) on the left; ($\alpha$=1.23,$\theta$=20$^o$), ($\alpha$=1.38,$\theta$=65$^o$), and ($\alpha$=1.45,$\theta$=35$^o$) while $P_x$=673nm is kept constant. (c)Simulated device transmission normalized with respect to the transmission through SiO$_2$/Si substrate for the 3 layer supercell simulation with the parameters given in b respectively. (d)Measured device transmission normalized with respect to the measured transmission through the SiO$_2$/Si substrate, for the devices simulated in c. (e)The norm of the electric field for the resonances given in c, the largest wavelength resonators sit at the bottom layer and the smallest wavelength resonators sit at the top layer.}
\label{fig:fig_3layer_3wvl}
\end{figure}

\begin{figure}
\centering
\includegraphics[width=5 in]{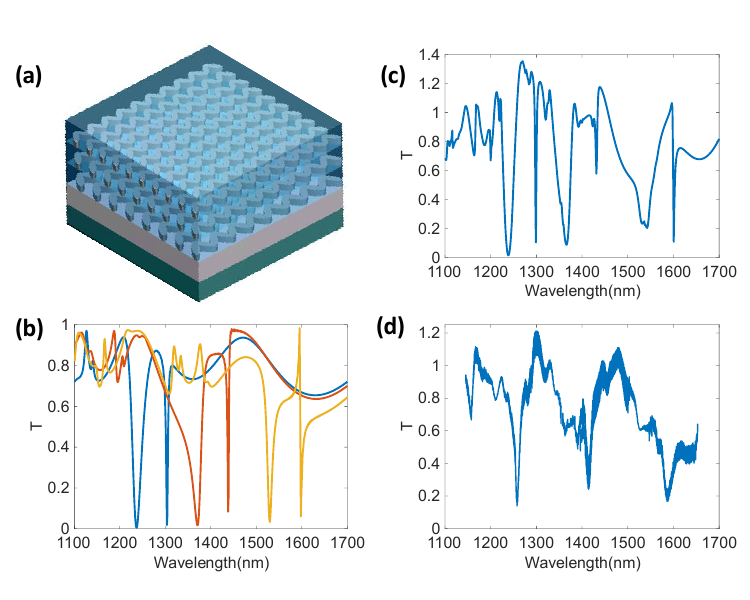}
\caption{(a)The supercell for the square lattice 3 layer device simulation. (b)Single layer simulations for ($P_y$=$P_x$=802nm,$\theta$=35$^o$),($P_y$=$P_x$=893nm,$\theta$=35$^o$), and ($P_y$=$P_x$=1002nm,$\theta$=35$^o$). (c)Simulated device transmission normalized with respect to the transmission through SiO$_2$/Si substrate for the 3 layer supercell simulation with the parameters given in b. (d)Measured device transmission normalized with respect to the measured transmission through the SiO$_2$/Si substrate, for the device simulated in c.}
\label{fig:fig_3layer_6wvl}
\end{figure}

Three-layer device simulations are performed using structures in which each layer has a different lattice constant $P_y$, while $P_x$ is kept fixed for computational simplicity. Consequently, for each three-layer device simulation, a periodic supercell is constructed in which different unit-cell repetition counts are assigned to each layer. These repetition counts are chosen to introduce only a small deviation from the target lattice constant $P_y$ per each layer (deviation less than 0.6\%), thereby enabling the use of periodic boundary conditions in the finite-difference time-domain (FDTD) simulations, as illustrated in Figure \ref{fig:fig_3layer_3wvl}.a (see details in Optical Design with FDTD in Supporting Information). The primary design parameters in this framework are the lattice-scaling parameter $\alpha$ and the post rotation angle $\theta$, as defined earlier. Using this design space, we design and experimentally demonstrate two representative devices to illustrate the full range of achievable spectral tuning. Device 1 (Figure \ref{fig:fig_3layer_3wvl}.b, left) serves as a reference design, featuring uniform resonance wavelength spacing and similar quality factors across all layers. In device 2 (Figure \ref{fig:fig_3layer_3wvl}.b, right), selective parameter variations are introduced: the rotation angle $\theta$ is modified only in the first (top) layer, both $\theta$ and $\alpha$ are adjusted in the second (middle) layer, and the third (bottom) layer is kept identical to that of device 1. This layer-wise parameter variation demonstrates that the resonance wavelength and quality factor $Q$ can be tuned independently on a per-layer basis, while leaving the spectral response of the unmodified layers essentially unchanged. As will be shown, this result highlights the decoupled nature of the multilayer design and its suitability for constructing independently programmable spectral filters.

Device 1 consists of three layers in which $\alpha$ takes values of 1.23, 1.34, and 1.45 from the first (top) to the third (bottom) layer (fabricated device shown in Figure S1), while $\theta$ is kept at $35^\circ$ throughout the device. On the other hand, device 2 consists of three layers in which $\alpha$ is set to 1.23, 1.38, and 1.45, and $\theta$ is set to $20^\circ$, $65^\circ$, and $35^\circ$ from top to bottom. The parameters required for targeting specific wavelengths are calculated by first conducting single-layer simulations (Figure \ref{fig:fig_3layer_3wvl}.b).
The three-layer device simulations for device 1 exhibit resonance dips at 1224, 1321, and 1453 nm (Figure \ref{fig:fig_3layer_3wvl}.c, left), while the experimentally measured dips occur at 1247 nm (23 nm shift), 1364 nm (43 nm shift), and 1509 nm (56 nm shift), respectively (Figure \ref{fig:fig_3layer_3wvl}.d, left). For device 2, simulations show resonances at 1224, 1390, and 1455 nm (Figure \ref{fig:fig_3layer_3wvl}.c, right), whereas the experimentally observed dips occur at 1243 nm (19 nm shift), 1431 nm (41 nm shift), and 1512 nm (57 nm shift) (Figure \ref{fig:fig_3layer_3wvl}.d, right).
The 23 nm and 19 nm shifts of the first resonances can be attributed to fabrication discrepancies, as these values are reasonably close to the simulated wavelengths. In contrast, the wavelength shifts increase for the lower layers. This behavior can be explained by the bottom spacer layers being exposed multiple times as additional layers are added and patterned. We observe average redshifts of 21 nm and 35.5 nm for layers 1 and 2, respectively, when fabricated geometry discrepancies are taken into account. HSQ is an electron-sensitive material, and its refractive index changes upon electron exposure. As shown in Figure S3, a change in the HSQ refractive index of 0.045 results in a 21 nm shift, while a change of 0.067 results in a 35.5 nm shift. These values are reasonable when compared with the limited literature on refractive index changes in HSQ as a function of electron-beam dosage \cite{choi2008}.
Finally, we note that 1.4 $\mu$m thick HSQ spacers are used between layers, as this thickness was found to be the minimum required for the layers to remain sufficiently decoupled from one another. This is evidenced by the electric-field profiles shown in Figure \ref{fig:fig_3layer_3wvl}.e for the resonances in Figure \ref{fig:fig_3layer_3wvl}.c (left). The 1453 nm resonators reside in the bottom layer, the 1321 nm resonators in the middle layer, and the 1224 nm resonators in the top layer, as indicated in Figure \ref{fig:fig_3layer_3wvl}.e.

As mentioned above, we also designed a six-wavelength device. In this case, the repetition of the unit cells for each layer is performed along both the $x$ and $y$ directions due to the square lattice configuration (Figure \ref{fig:fig_3layer_6wvl}.a). As before, single-layer simulations are first conducted to determine the appropriate design parameters (Figure \ref{fig:fig_3layer_6wvl}.b). The three-layer device simulations exhibit inter-post resonance dips at 1239, 1370, and 1533 nm (Figure \ref{fig:fig_3layer_6wvl}.c), whereas experimentally the resonances appear at 1257 nm (18 nm shift), 1413 nm (43 nm shift), and 1588 nm (55 nm shift) (Figure \ref{fig:fig_3layer_6wvl}.d).
As before, the 18 nm shift of the first resonance can be attributed to geometric deformations during fabrication, while the larger relative shifts of 25 nm and 37 nm are comparable to those observed in the three-wavelength device and can be explained by refractive index changes in the HSQ spacer upon the addition and exposure of subsequent layers. In addition, higher-order hybrid modes appear as shallow dips in the transmission spectrum of the three-layer device. It can be observed that the six-wavelength design layers are both numerically and experimentally more decoupled than the three-wavelength design, as evidenced by the absence of additional resonances with coupling strengths comparable to those of the primary resonances. Compared to the three-wavelength design, the six-wavelength design displays a simpler transmission response with three dominant resonances, while additional coupling-induced features remain weaker. Although the high-$Q$ intra-post modes do not appear experimentally, this behavior results in a cleaner measured spectrum compared to the three-wavelength design. Consequently, depending on the target application, either design strategy may be employed to realize a multi-resonant filter with three operational resonance wavelengths. From a spectral construction perspective, as long as direct and independent control over the filter spectrum is maintained through the available design parameters ($\alpha$ and $\theta$), the presence of additional higher-order modes at distinct wavelengths can further enrich the spectral response and enhance compressive sensing performance.

\subsection{Filter Array for Compressive Sensing}

In addition to having diverse spectral features for compressive sensing, a second parameter needed for such reconstruction is to have a set of basis functions with minimum correlation. In spectral reconstruction, the goal is to estimate the incident optical spectrum from measurements obtained using a limited number of arrayed filters. Designing metasurface filters with decorrelated or minimally overlapping spectra reduces the mutual correlation of the measurement matrix, thereby stabilizing the inversion problem and improving noise robustness \cite{oliver2013,wang2025,guo2024,wang2014}. Moreover, low-correlation basis spectra increase the system’s ability to resolve narrow spectral features and reduce the number of measurements required for compressed sensing-based recovery. Owing to their capability for precise tailoring of transmission and reflection spectra, metasurfaces have emerged as particularly promising platforms for constructing such spectral basis sets \cite{hu2024,zhao2025,tan2025,he2024}.

By enabling fine control over resonance wavelengths and Q-factors across multiple layers, our multi-resonant metasurfaces naturally generate a set of decorrelated spectral transfer functions that serve as an advantageous basis for stable and high-fidelity spectral reconstruction. Each multilayer filter exhibits a rich spectral response, combining both broadband and narrowband features arising from resonances localized in individual layers, as well as from higher-order hybrid modes. A key advantage of this platform is the direct and independent control over the shape of each basis spectrum through deterministic tuning of the resonance wavelengths and their associated $Q$ factors. Previous approaches have demonstrated the use of single-layer photonic crystal slabs \cite{wang2025} and free-form inverse-designed metasurfaces \cite{chen2023,chen2025,hu2024} for spectral reconstruction. Multilayer filter architectures have also been proposed to further reduce mutual correlations between basis spectra; however, these implementations typically rely on stacks of thin films or combinations of metasurface pillars and thin-film layers to achieve spectral diversity \cite{liu2021,yao2025,guo2024,wang2014}. Such architectures are generally not well suited for single-chip operation, as different filter regions require different overall thicknesses.
In contrast, the approach presented here maintains identical spacer thicknesses across all filters and achieves spectral tuning exclusively through geometric modification of the elliptical resonators. This uniform-thickness design is inherently compatible with single-chip integration.

\begin{figure}
\centering
\includegraphics[width=5 in]{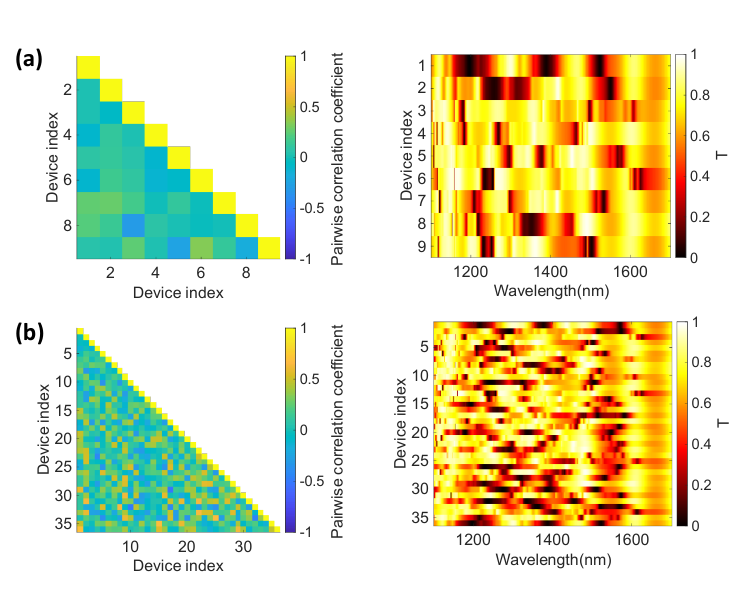}
\caption{(a)Pairwise correlation coefficients for the selected 9 devices with average absolute value of the pairwise correlation coefficients 0.11 (left) and the heatmap for the transmission spectra of these filters (right). (b)Pairwise correlation coefficients for the selected 36 device with average absolute value of the pairwise correlation coefficients 0.21 (left) and the heatmap for the transmission spectra of these filters (right).}
\label{fig:fig_corr}
\end{figure}

In literature, the correlation coefficient of such basis functions has been used as the main metric to characterize and justify such spectra \cite{wang2025,hu2024,chen2023,chen2025}. Here we use Pearson correlation coefficient to calculate the extent to which the spectrum set is linearly independent using the expression:

\begin{equation}
r_{xy} = 
\frac{
\sum_{i=1}^{n} (x_i - \bar{x})(y_i - \bar{y})
}{
\sqrt{\sum_{i=1}^{n} (x_i - \bar{x})^{2}}
\,
\sqrt{\sum_{i=1}^{n} (y_i - \bar{y})^{2}}
}
\end{equation}

where $x_i$($y_i$) is the transmission value at wavelength $\lambda_i$ for arbitrary two filters, and $\bar{x}$($\bar{y}$) is the mean of $x_i$($y_i$) values for i=1,2,..,601 (from 1100nm to 1700nm).
Changing the parameters $\alpha$ from 1.147 to 1.587 (17 points) and $\theta$ from 5$^o$ to 80$^o$ (6 points), we chose the set of filters  shown in Fig. \ref{fig:fig_corr}. Again, to be able to operate on a single chip, spacer thicknesses are kept the same (1.4 um) throughout the filter array. 

In order to do an extensive literature comparison, we generate a set of 9 and a set of 36 filters with average absolute values of the pairwise correlation coefficients $r_{xy}$ 0.11 (Fig. \ref{fig:fig_corr}.a) and 0.21 (Fig. \ref{fig:fig_corr}.b) respectively in the 1100-1700 nm band. To efficiently design a diverse set of multi-resonant filters, we first approximate three-resonance transmission spectra using 102 single-layer simulations (17 values of $\alpha$ and 6 values of $\theta$). For every combination of three single-layer resonances, an estimated multilayer response was constructed by taking the point-wise minimum of the corresponding single-layer transmission curves, which provides a fast surrogate for the composite three-layer device spectrum. The Pearson correlation coefficients between all estimated spectra were then computed, and a greedy max–min diversity selection algorithm \cite{gonzalez1985} was applied to extract subsets of 9 and 36 candidate parameter sets that minimize pairwise spectral similarity. In this algorithm, the first spectrum is chosen based on its overall variance, and subsequent spectra are added iteratively by selecting, at each step, the candidate whose maximum absolute correlation with all previously selected spectra is the smallest. This ensures that each newly added spectrum contributes maximal additional diversity to the selected set. The resulting set of 9 and 36 filters were subsequently validated through full 3D simulations of the actual three-layer devices, confirming that the greedy selection procedure reliably identifies parameter sets that yield low-correlated, spectrally distinct filter responses. When we compare the average correlation values with the literature examples mentioned above, we see that Wang et \textit{al.} design a 3x3 array with average correlation coefficient 0.17 with photonic crystal slabs for 8-11.5 um wavelength band \cite{wang2025}; Hu et \textit{al.} design a 5x5 array with average correlation coefficient 0.22 with freeform metasurfaces for 380nm-680nm wavelength band \cite{hu2024}; Chen et \textit{al.} design a 7x7 array with average correlation coefficient 0.41 with freeform metasurfaces for 400nm-700nm wavelength band \cite{chen2023}; and, Chen et \textit{al.} design a 6x6 array with average correlation coefficient 0.49 with freeform metasurfaces for 500nm-700nm wavelength band \cite{chen2025}. Our filters (both set of 9 and 36) surpass these studies in terms of decorrelation within a comparable operation band.

\section{Conclusion}
In conclusion, we have demonstrated a high-throughput direct-write fabrication platform based on an antimony–BDCA precursor for multilayer resonant metasurfaces. This work greatly expands the design space of multilayer devices that has remained comparatively unexplored due to the complexity and low yield of standard nanofabrication flows. By combining Sb$_2$S$_3$ q-BIC–derived metasurfaces with HSQ planarization layers, we realize three-layer architectures in which each layer remains largely decoupled, enabling independent control of multiple resonances through the geometric parameters $\alpha$ and $\theta$ while keeping all spacer thicknesses identical across the chip. This yields compact multi-resonant filters whose transmission spectra can be arbitrarily tailored at design time, providing rich combinations of broad and narrow features that are directly encoded by device geometry rather than by varying film thickness. We leveraged this capability to generate decorrelated filter arrays that are well suited for spectral reconstruction based on compressive sensing, demonstrating sets of 9 and 36 filters with average absolute pairwise Pearson correlation coefficients of $|r_{xy}| = 0.11$ and $0.21$, respectively, surpassing previously reported metasurface and photonic-crystal–based implementations in both decorrelation and operational bandwidth. Beyond the specific 3 and 6 wavelength devices realized here, this platform paves the way towards structures with more patterned layers to enable densely-spaced resonances and integration with on-chip detector arrays. Ultimately, we envision application-specific metasurface filters where arbitrary low-correlated spectral basis arrays can be co-designed and monolithically fabricated for compact, high-performance hyperspectral imaging and compressive spectroscopy systems.


\medskip
\textbf{Acknowledgements} \par 
This research was mainly supported by the Army Research Office (W911NF-22-1-0097), Caltech Sensing to Intelligence, and Carver Mead New Adventures Grant. Research was in part carried out at the Molecular Materials Research Center in the Beckman Institute of the California Institute of Technology. We thank Prof. Julia Greer for letting us use the Schlenk line in their lab for the precursor synthesis. We also thank Dr. Wei Wang for fruitful discussions regarding the precursor synthesis and the EBL direct patterning discussed in Ref. \cite{Wang2020}. The devices are cleaved for microscopy using SELA MC20 micro cleaving system. We thank Ian Foo and Suki Gu for the discussions regarding design and fabrication. 

\medskip
\textbf{Conﬂict of Interest} \par 
The authors declare no conflicts of interest.

\medskip
\textbf{Data Availability Statement} \par
Data underlying the results presented in this paper are not publicly available at this time but may be obtained from the authors upon reasonable request.

\medskip



\clearpage

\setcounter{section}{0}
\setcounter{figure}{0}
\setcounter{table}{0}









\addcontentsline{toc}{section}{Supporting Information}

\begin{center}
\textbf{\LARGE Supporting Information}\\[0.5em]
\textbf{Multilayer Q-BIC-like Optical Filters with High Throughput Direct-Write Multilayer Lithography}\\[1em]
\end{center}

\bigskip

\section{Fabrication Details}

\subsection{Precursor Solution Synthesis}  
The precursor was synthesized by reacting 1-butylamine (2 mL) with carbon disulfide (1.5 mL) to get BDCA (reaction 1). The precursor, which is a Sb-BDCA  complex is created mixing antimony (III) oxide (1 mmol) dissolved in ethanol (2 mL), and BDCA (reaction 2). When this complex is exposed to an electron bombardment, it decomposes into Sb$_2$S$_3$ in addition to other organic byproducts (reaction 3). Unexposed area and the byproducts of the exposed area remain organic and washed away by immersing in isopropanol (IPA) for 10 mins.

\begin{figure}[h]
\centering
\includegraphics[width= 6 in]{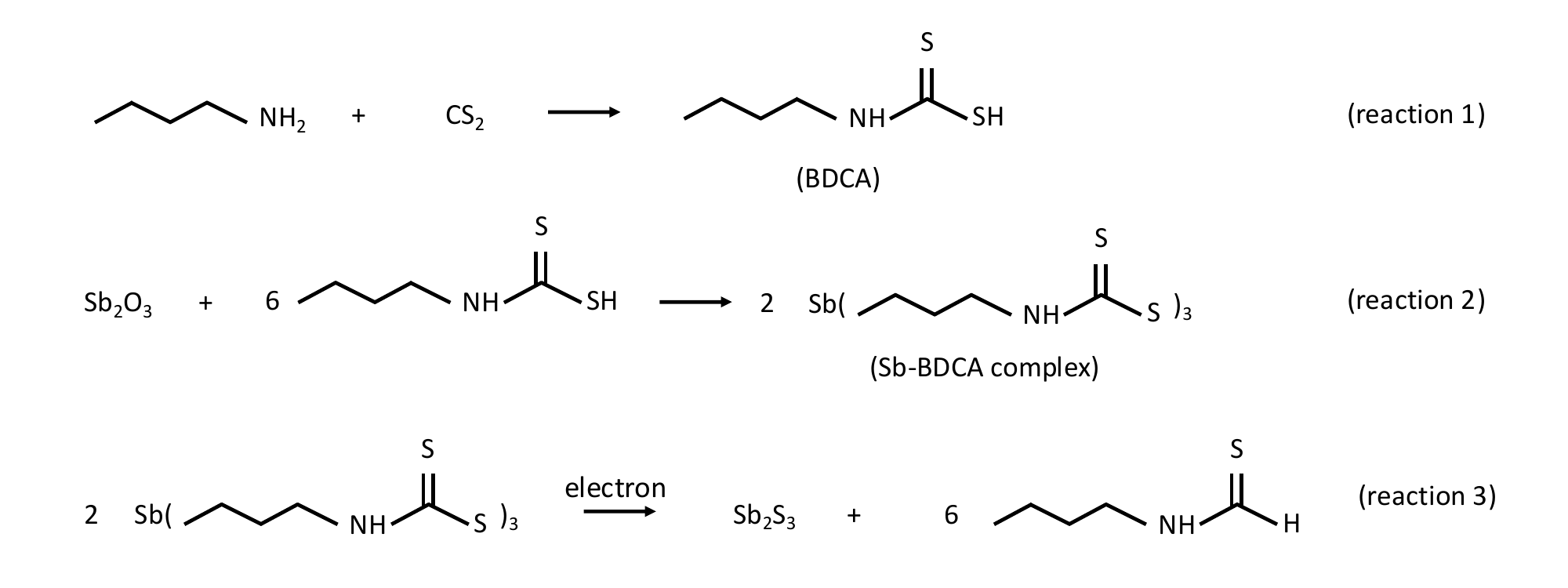}
\label{fig:reactions}
\end{figure}

The synthesis procedure follows: In a round-bottom flask, antimony (III) oxide is dissolved in ethanol and carbon disulfide at room temperature as reported in Ref. \cite{Wang2020}. The flask was connected to a Schlenk line (nitrogen flow, no vacuum applied) through a glass condenser, and the mixture was stirred under nitrogen. Subsequently, 1-butylamine was added to the flask drop-by-drop since the reaction is exothermic, after which the reaction was allowed to proceed under continuous stirring overnight.

\subsection{Preparation of the Substrate}
1300 nm of silica was deposited using PECVD on a double sided polished (DSP) doped silicon wafer. We fabricated gold markers on silica deposited silicon chips by patterning an electron beam resist ZEP520A using EBL, depositing 50 nm gold with electron beam deposition, and lifting off the excess gold. We used these markers sitting on the substrate for all of the layers patterned and no additional markers were fabricated per added layer.

\begin{figure}[h]
\centering
\includegraphics[width=6 in]{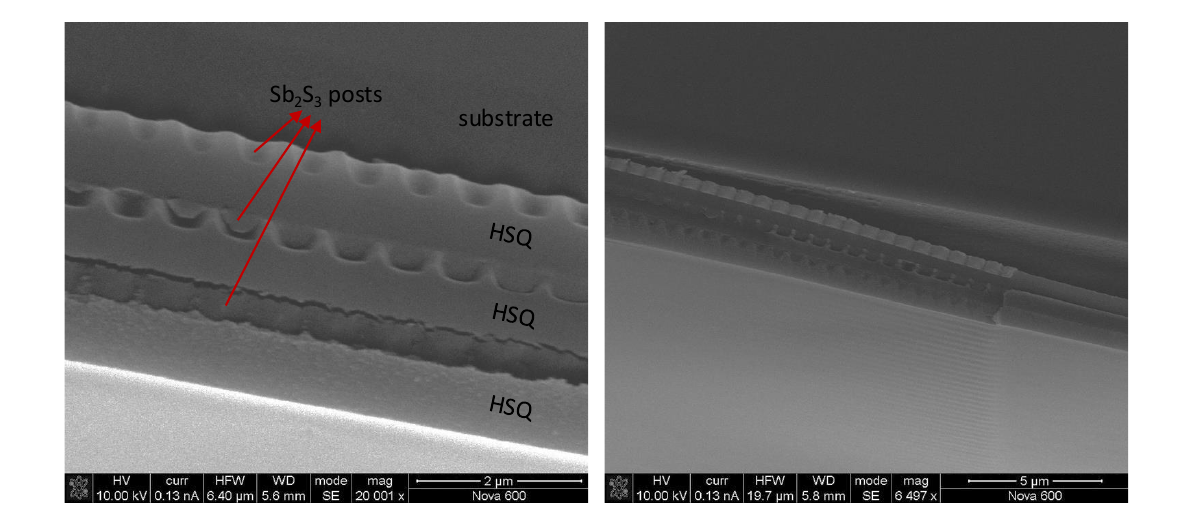}
\caption*{Figure S1: SEM image of the cleaved 3 layer device given in main manuscript Figure 4.b and c. on the left, zoomed out cross section SEM where the posts are still intact on the right (edge lifted during cleaving.}
\label{fig:sem}
\end{figure}

\subsection{EBL Patterning} 
For film deposition, 1 mL of the precursor solution was diluted with 2 mL of ethanol and spin-coated at 6500 rpm to form a uniform thin film. This spin-coating step is repeated once more before exposure to have a thicker spin-coated film. The coated substrates were then patterned using a 100 kV EBL system with a beam current of 20 nA. A multipass exposure strategy was employed, consisting of 10 passes with a dose of 50 mC/cm² per pass (total accumulated dose of 500 mC/cm²). A multipass approach is preferred to avoid dumping all the exposure energy at a single location and then switch to the next writing grid. Instead energy exposure is done bit by bit over all patterned region. This makes the resulting patterns much closer to the intended write patterns. For continuous pattern that are larger than the subfield of the EBL tool (4 um), this becomes a neccessity to avoid cracks.

Following the exposure, the samples were developed in IPA, and subsequently annealed at 180 °C for 2 min on a hotplate inside a nitrogen glovebox. For planarization, HSQ (FOx-16) was spun at 5000 rpm for 60s twice to reach to a thickness of approximately 1.4 um on the Sb2S3 structures. Here, HSQ was used only as a spin-on-glass for planarization (not as an electron beam sensitive medium). The spun HSQ was annealed at 180 °C for 2 mins for partial curing. These steps are repeated to fabricate each layer. The 3 layer device mentioned in the manuscript can be seen in Figure S1, left. Here, since $P_y$ in each layer is different, the cross section of the ziczac pattern appears as post cross sections with different sizes. We note that the posts in the first and second layer fell due to cleaving, however, we also maned to image regions of the device where the posts were still intact (Figure S1, right).

\section{Chemical and Optical Characterization}

\begin{figure}[h]
\centering
\includegraphics[width=\linewidth]{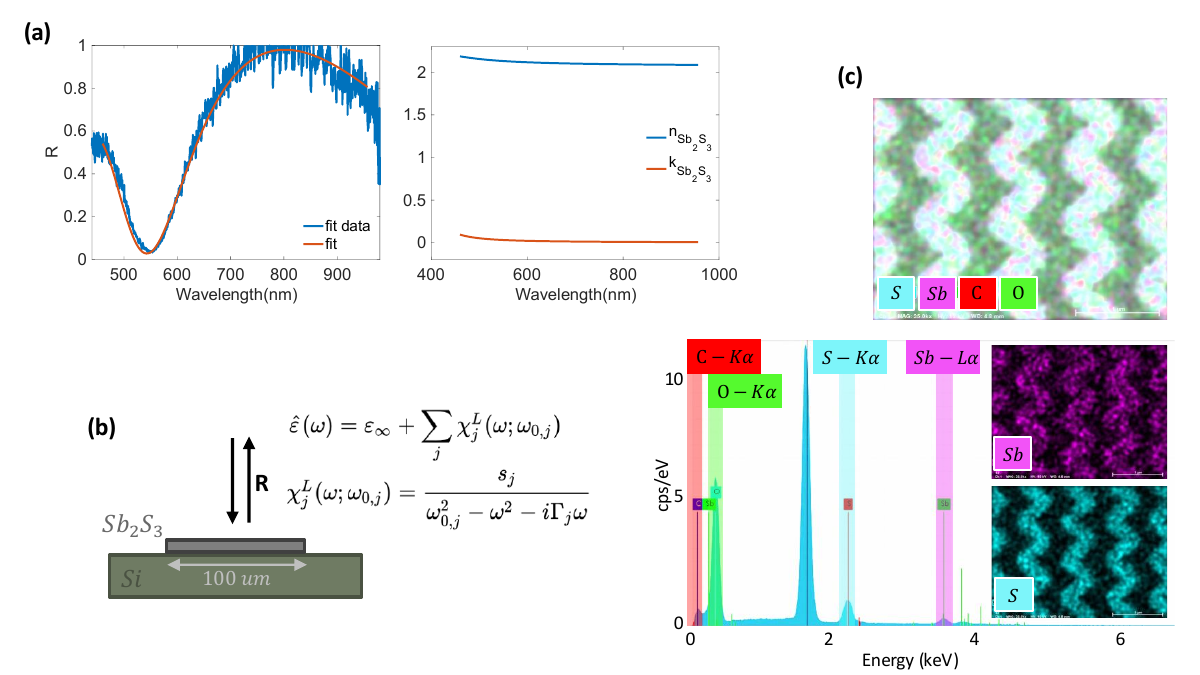}
\caption*{Figure S2: (a)Measured reflection from the sample and the fit with the oscillator model (left) and the consequent $n,k$ data (right). (b)100umx100um large Sb$_2$S$_3$ pattern on Si substrate for the reflection measurements and the Lorentz oscillator model used where $\varepsilon_\infty$ is a parameter that captures the effects of higher energy contributions to the dielectric function, and $s$, $\omega_0$, $\omega$, $\Gamma$ are the oscillator strength, resonant frequency,
optical frequency, and oscillator linewidth, respectively. (c)EDS measurement showing the chemical composition of the nanostructures patterned on the substrate where cyan denotes sulfur (S), magenta denotes antimony (Sb), red denotes carbon (C), and green denotes oxygen (O).}
\label{fig:fig_char}
\end{figure}

Although there are several reports on the refractive index of Sb$_2$S$_3$ ($n$), the value of $n$ depends strongly on the synthesis conditions of the precursor as well as subsequent processing parameters such as precursor dilution during spin coating and the phase of the resulting material \cite{sb2s3refindex_delaney2020,gutierrez2022,2020}.

Direct determination of the refractive index using ellipsometry is impractical in this case, as it would require patterning a continuous, centimeter-scale Sb$_2$S$_3$ film, which is prohibitively time-consuming due to the high electron beam doses involved. Instead, the dielectric function of Sb$_2$S$_3$ is extracted by fitting reflection measurements performed on a thin Sb$_2$S$_3$ film using the transfer matrix method (TMM) (Figure S2.a).

To this end, we fabricated continuous $100~\mu$m $\times$ $100~\mu$m square Sb$_2$S$_3$ patterns on a silicon substrate and performed reflection measurements using a supercontinuum source coupled to a spectrometer over the wavelength range of 450–950 nm. The incident beam was focused to a spot size of approximately $\sim 70~\mu$m on the Sb$_2$S$_3$ film. The measured reflectance spectrum was fit using a thin-film interference model based on TMM \cite{yariv_photonics}, enabling extraction of both the real ($n$) and imaginary ($k$) components of the refractive index. Specifically, the dispersive response of the film is parameterized using a Lorentz oscillator model, with the functional form shown in Figure S2.b. Model fitting is performed using a particle swarm optimization (PSO) algorithm. The fitting wavelength range is chosen to be 450–950 nm, rather than the device operation band, as the reflectance data in this region exhibits stronger spectral features that enable more reliable extraction of the dielectric function. Given the monotonic behavior of the extracted refractive index, a constant value of $n = 2.1$ is assumed over the device operation band, which will be shown to agree with the experimental results.

The chemical composition of the patterned Sb$_2$S$_3$ structures is further confirmed using energy-dispersive X-ray spectroscopy (EDS). As shown in Figure S2.c, the EDS spectra clearly indicate the presence of antimony and sulfur, confirming the chemical decomposition of the precursor into Sb$_2$S$_3$. Additional signals corresponding to silicon, oxygen, and carbon arise from the substrate, the underlying spacer layer, and the conductive coating used for scanning electron microscopy (SEM), respectively.

\section{Optical Design with FDTD}
For single layer simulations two rotated Sb$_2$S$_3$ elliptical posts (n=2.1) that are 400 nm thick (rotated by $\theta$) are placed on a 1.3 um thick SiO$_2$ bottom spacer on a Si substrate. Here, the lattice constants are set as $P_x = s*1020$ nm and $P_y = \alpha * P_x$. The length of the semi-major axes of the elliptical posts are set as $s*165$ nm and $s*500$ nm respectively where s is the scaling factor for all lateral features. This constant scaling factor s is applied to all lateral features including the ellipses and the lattice to make Q of the resonance only depend on the tilt angle $\theta$. In Figure 3.b, for the square lattice simulation, s=0.870 and $\alpha$=1, and for the rectangular lattice simulation shown in Figure 3.c, s=0.660 and $\alpha$=1.34. The planarization layer is set to HSQ with $n$=1.4. The top and the bottom boundary conditions are set as perfectly matched layer (PML) where the side surfaces are set to Bloch periodic boundary conditions. As for the 3 layer simulations, for each layer $i$ with nominal lattice constant $P_{y,i}$, we choose an integer repetition number $N_i$ and define a common supercell length $L_y$ such that $L_y = N_i P'_{y,i}$, where $P'_{y,i}$ is a slightly adjusted lattice constant. Because the nominal periods $P_{y,i}$ generally do not form an exact integer ratio, we select $\{N_i\}$ so that the adjusted values $P'_{y,i} = L_y / N_i$ deviate minimally from the design values (less than 0.6\%). 

\subsection{HSQ Refractive Index Change}
The shift between the simulated resonance wavelength and what is measured can be explained by the index change of HSQ with the exposure per added layer as shown in Figure S3.

\begin{figure}[h]
\centering
\includegraphics[width=6 in]{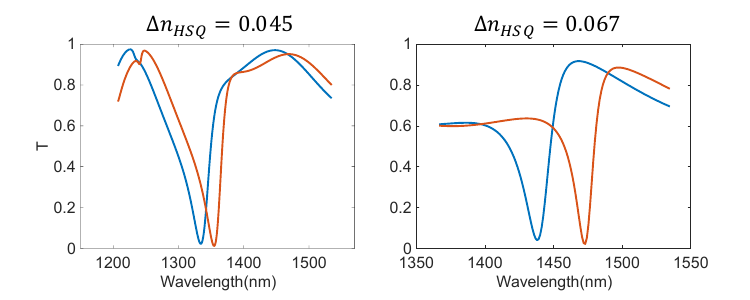}
\caption*{Figure S3: The shift of the resonance with the change in the HSQ refractive index $\Delta n_{HSQ}$=0.045, and $\Delta n_{HSQ}$=0.067. }
\label{fig:sem}
\end{figure}

\bibliographystyle{MSP}
\bibliography{manuscript_ref_v4}
\end{document}